\documentclass[12pt]{article} 
\usepackage[sectionbib]{natbib}
\usepackage{array,epsfig,fancyheadings,rotating}
\usepackage{amsmath}
\usepackage{url}
\usepackage[]{hyperref}  
\usepackage{sectsty, secdot}
\sectionfont{\fontsize{12}{14pt plus.8pt minus .6pt}\selectfont}
\renewcommand{\theequation}{\thesection\arabic{equation}}
\subsectionfont{\fontsize{12}{14pt plus.8pt minus .6pt}\selectfont}

\usepackage{amsmath}
\usepackage{amssymb}
\usepackage{amsfonts}
\usepackage{multirow}
\usepackage{amsthm}
\usepackage{color}

\newtheorem{theorem}{Theorem}

\newtheorem{proposition}{Proposition}
\theoremstyle{definition}

\begin{document}
	
	
	\renewcommand{\baselinestretch}{2}
	
	\markright{ \hbox{\footnotesize\rm Statistica Sinica
		}\hfill\\[-13pt]
		\hbox{\footnotesize\rm
		}\hfill }
	
	\markboth{\hfill{\footnotesize\rm QIUYU WU AND XIANGYU LUO} \hfill}
	{\hfill {\footnotesize\rm Nonparametric Bayes for two-level clustering} \hfill}
	
	\renewcommand{\thefootnote}{}
	$\ $\par
	
	
	\fontsize{12}{14pt plus.8pt minus .6pt}\selectfont \vspace{0.8pc}
	\centerline{\large\bf Nonparametric Bayesian Two-Level Clustering}
	\vspace{2pt}
	\centerline{\large\bf for Subject-Level Single-Cell Expression Data}
	\vspace{.4cm}
	\centerline{Qiuyu Wu, Xiangyu Luo
	}
	\vspace{.4cm}
	\centerline{\it Institute of Statistics and Big Data, Renmin University of China}
	\vspace{.55cm} \fontsize{9}{11.5pt plus.8pt minus.6pt}\selectfont
	
	
	\begin{quotation}
		\noindent {\it Abstract:} The advent of single-cell sequencing opens new avenues for personalized treatment. In this paper, we address a {\it two-level clustering} problem of simultaneous subject subgroup discovery ({\it subject level}) and cell type detection ({\it cell level}) for single-cell expression data from multiple subjects. However, current statistical approaches either cluster cells without considering the subject heterogeneity or group subjects without using the single-cell information. To bridge the gap between cell clustering and subject grouping, we develop a nonparametric Bayesian model, Subject and Cell clustering for Single-Cell expression data (SCSC) model, to achieve subject and cell grouping simultaneously. SCSC does not need to prespecify the subject subgroup number or the cell type number. It automatically induces subject subgroup structures and matches cell types across subjects. Moreover, it directly models the single-cell raw count data by deliberately considering the data's dropouts, library sizes, and over-dispersion. A blocked Gibbs sampler is proposed for the posterior inference. Simulation studies and the application to a multi-subject iPSC scRNA-seq dataset validate the ability of SCSC to simultaneously cluster subjects and cells.

		\vspace{9pt}
		\par
	\end{quotation}\par

	\def\thefigure{\arabic{figure}}
	\def\thetable{\arabic{table}}
	
	\renewcommand{\theequation}{\thesection.\arabic{equation}}

	\fontsize{12}{14pt plus.8pt minus .6pt}\selectfont

	\section{Introduction}
\begin{figure}
		\centering
		\includegraphics[scale = 0.4]{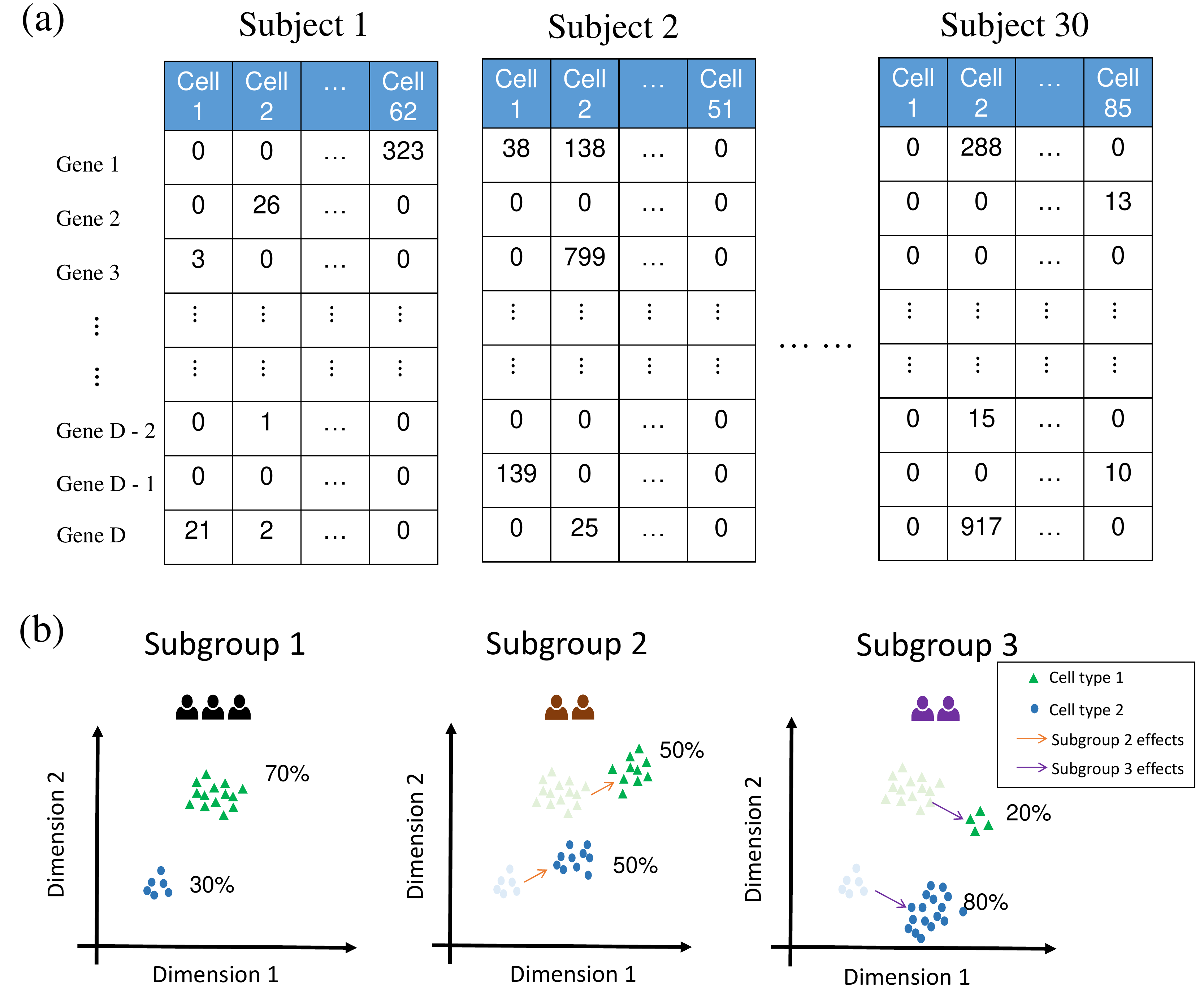}
		\caption{Artificial illustration of the data structure and study goal. (a) Subject-level single-cell expression data. (b) The illustration of a two-level clustering problem. In subgroup 1, cell type 1 is $70\%$ in green triangles, and cell type 2 is $30\%$ in blue dots. Compared to subgroup 1, the cellular distribution in subgroup 2 can change in two ways: cell proportions and cell locations. For a good visualization, only two gene dimensions are illustrated (expression in log scale). The orange and purple arrows represent subgroups 2 and 3 effects, respectively, when subgroup 1 is treated as a reference.}\label{fig1}
	\end{figure}

	Advancements in biological sequencing technology, such as single-cell RNA-sequencing (scRNA-seq), have enabled the expression profiling of single cells. ScRNA-seq data are often organized into a data matrix illustrated in Figure \ref{fig1}(a), where columns are cells and rows represent genes. Based on the scRNA-seq data matrix, discovering cell types is simply formulated as a clustering problem. Going further, if we can integrate the scRNA-seq data from multiple subjects, it presents unprecedented opportunities to investigate subject heterogeneity at the single-cell resolution. Subject heterogeneity refers to human subpopulations, patient disease subtypes, or other differentiable human biological characteristics according to different contexts. Using the disease subtypes as an illustration, biological studies have found differences in tumor cell proportions among subtypes of breast cancers \citep{Makki2015Diversity}, lung cancers \citep{busch2016lung}, and other diseases. The subtle observations can be captured by the scRNA-seq data but may be missed using the traditional bulk expression data, which are the aggregated expression signals from diverse cell types. Consequently, it is imperative to employ the subject-level single-expression data (Figure \ref{fig1}(a)) to understand cellular and subject heterogeneity.
	
	In this study, we aim to address a two-level clustering statistical problem by directly modeling the multi-subject scRNA-seq data. An artificial demonstration of the two-level clustering is shown in Figure \ref{fig1}(b). At the cell level, the cells having similar expression values are clustered together and at the subject level, the subjects having similar cellular distributions are grouped together. Two subjects are said to have the same cellular distributions if they share the same cell type proportions and expression levels for each cell type. In addition, to obtain valid biological results, cell types must be matched across subjects by considering the effects caused by the subject subgroups (Figure \ref{fig1}(b)). We notice that our two-level clustering problem is different from the bi-clustering approaches \citep{cheng2000biclustering, turner2005improved}, which group subjects and genes using the aggregated expression data matrix.
	
	There has been a large amount of statistical literature on cell clustering or subject clustering. On one hand, cell clustering methods fit the heterogeneous scRNA-seq data via latent variable model \citep{buettner2015computational}, hierarchical clustering \citep{yau2016pcareduce}, consensus approach \citep{Kiselev2017SC3}, or model-based mixture models \citep{Prabhakaran2016Dirichlet, Sun2017DIMM, song2020flexible, liu2019hierarchical}. Nevertheless, when applied to multi-subject scRNA-seq data, these methods do not consider the subject heterogeneity, and they ignore the fact that the gene expression levels may change with subjects, thus possibly leading to incorrect cell clustering results.
	
	On the other hand, subject clustering methods are based on the aggregated expression matrix with genes in rows and subjects in columns, where the expression vector of one subject can be viewed as the row averages of the subject's gene-cell expression matrix in Figure \ref{fig1}(a). \citet{pan2007penalized} adopted a normal mixture model and developed an $L_1$-penalized expectation-maximization algorithm to distinguish subjects and detect differentially expressed (DE) genes. \citet{wang2008variable} instead used the $L_{\infty}$ and hierarchical penalties to refine the clustering results. The sparse k-means proposed by \citet{witten2010framework} simultaneously extracted a few DE genes and grouped subjects by maximizing the weighted between-cluster sum-of-squares. \citet{huo2016meta} subsequently generalized sparse k-means to expression data from multiple studies. \citet{luo2019batch} proposed a more efficient and flexible Bayesian framework to conduct integrative subject clustering. Since these methods do not employ single-cell expression information, subtle differences (e.g., cellular composition changes) cannot be detected.
	
	All the methods mentioned above except \citet{Prabhakaran2016Dirichlet} require predetermination of the number of clusters and the trial of multiple choices, which may be practically difficult and computationally expensive. The Dirichlet process is a nonparametric Bayesian prior \citep{ferguson1973a, Sethuraman1991A} and is well-known for its flexibility in automatically selecting the number of clusters in a data-driven manner. However, the DP only addresses one-level clustering, motivating two extensions---the hierarchical DP (HDP) \citep{hdp} and the nested DP (NDP) \citep{rodriguez2008nested}---that are close to our two-level clustering problem. Unfortunately, using the terms in our context, the HDP assigns a cell mixture distribution to each subject but with different mixture weights; thus, the subjects cannot form a group structure. Although the NDP promotes the subject group structure, subjects in different groups do not share any cell components, causing difficulty in matching cell types across subjects. In other words, if two distributions from the NDP share one cell component, the two distributions must be the same almost surely, which is not realistic in our problem. To deal with the degeneracy issue of the NDP,  \citet{camerlenghi2019latent} developed a latent nested nonparametric prior which allows common and group-specific cell types across subject subgroups, but their method meets practical computational challenges when applied to more than two subject subgroups or high-dimensional expression data. When more than two subject subgroups need to be considered, \citet{beraha2020semi} extended the HDP to semi-HDP to induce subject dependence and grouped distributions using a finite-dimensional distribution over cluster indicators.

	Actually, in the discussion of the NDP paper \citep{rodriguez2008nested}, \citet{james2008discussion} has constructed a fully nonparametric prior to combine the NDP and the HDP, which can address the degeneracy problem of the NDP and achieve two-level clustering for nested data. We follow the section 4 name in his discussion and call his prior {\it hybrid NDP-HDP prior}. In the filed of text analysis, the hybrid prior has been employed to conduct entity-topic modeling \citep{tekumalla2015nested}, and its multi-level extension introduced in \citep{paisley2014nested} allows tree-structured topic hierarchies. Recently, \citet{denti2020common} proposed a common atoms model built upon a similar nonparamemtric prior to analyze the microbiome data that does not introduce an additional HDP part but constrains the common atoms of sampled distributions.
	
	To the best of our knowledge, there is no statistical approach to simultaneously tackle subject and cell clustering on the multi-subject scRNA-seq data. For the two-level clustering part, we took advantage of the hybrid NDP-HDP prior \citep{james2008discussion}, inducing shared components for cells and group structures for subjects. For the data modeling part, we exploited the zero-inflated Poisson-log-normal (ZIPLN) distribution with a Probit dropout mechanism, which accounts for the zero-inflation, over-dispersion, and count nature of scRNA-seq data. Integrating the nonparametric Bayesian prior with the ZIPLN distribution results in the proposed model, Subject and Cell clustering for Single-Cell expression data (SCSC) model, which enables simultaneous subject and cell clustering for the scRNA-seq raw count data and does not require any specification for the subject or cell cluster number in advance. For the posterior inference of SCSC, we designed an efficient blocked Gibbs sampler \citep{ishwaran2001gibbs} based on an approximation to the SCSC model. The approximation accuracy is guaranteed theoretically as long as the truncation levels and related parameters are appropriately chosen.
	
	This paper is subsequently organized as follows. Section 2 presents a brief review of the DP and its two extensions, the HDP and the NDP, which are prerequisites to introduce the hybrid NDP-HDP prior that enjoys the strengths of the HDP and the NDP. In Section 3, we bring in the hybrid NDP-HDP prior, derive theoretical results about the distributions sampled from the prior, and present the SCSC model that is built on the hybrid prior and tailored to the scRNA-seq data. In Section 4, we introduce the truncated SCSC model to ease the posterior computing and provide a theorem to quantify its approximation error. An efficient posterior sampling scheme for SCSC is discussed in Section 5, and its application to synthetic and real-world data is illustrated in Section 6. Finally, we conclude the paper with a discussion in Section 7.	
	
	\section{Preliminaries on nonparametric priors}
	Suppose that the scRNA-seq data are collected for $m$ subjects with subject $j$ having $n_j$ sequenced cells in some tissue, and in each cell, the expression levels for $D$ genes are measured. We denote the observed read count mapped to gene $g$ in cell $i$ for subject $j$ by $X^{(j)}_{gi}$. All the read counts for subject $j$ can be wrapped up using a data matrix $\mathbf{X}^{(j)}$ with $D$ genes in rows and $n_j$ cells in columns. To describe the subject heterogeneity, we assume that subjects can be separated to form several subgroups, where subjects in the same subgroup share similar characteristics, and subjects in different subgroups have distinct features. We use $S^{(j)}$ to represent the subgroup which subject $j$ belongs to. Similarly, the cell heterogeneity is characterized by cell types, and the cell type of cell $i$ for subject $j$ is denoted by $C^{(j)}_i$. Note that $\mathbf{X}^{(j)}$'s are observed, but the subject subgroup and cell type indicators must be estimated.
	
	\subsection{Dirichlet process}
	The DP mixture model \citep{lo1984class} based on the DP prior \citep{ferguson1973a} can be considered as a generalized version of the finite mixture model. For notational simplicity, we temporarily consider only the cell data from subject $1$ and let the gene number $D$ be one. Thus, the column vectors $\mathbf{X}^{(1)}_1, \ldots, \mathbf{X}^{(1)}_{n_1}$ of $\mathbf{X}^{(1)}$ can be simplified to univariate samples $X_1, \ldots, X_{n_1}$ and the cell type indicators $C^{(1)}_i$'s to $C_i$'s. The finite mixture model allocates each cell to one of $K$ cell types with the probability of cell type $k$ being $\pi_k$, i.e., $\mathbb{P}(C_i=k)=\pi_k$, and $\sum_{k=1}^K\pi_k=1$. Given that cell $i$ is assigned to cell type $k$, $X_i$ is assumed to be from the distribution $f(x|\mu_k)$, where $f$ is a probability density (or mass) function, which will be specified in the next section, and $\mu_{k}$ is a parameter describing the cell-type-$k$ effect. Usually, the total cell type number $K$ is unknown to data analysts, and it is challenging to accurately estimate its value. The DP mixture overcomes this challenge by generalizing $K$ to infinity and allowing finite non-empty components, thereby not requiring a prespecification of $K$.
	
	The construction of the DP is realized by the stick-breaking process \citep{Sethuraman1991A}. Imagine that we have a stick of length 1 unit, and we intend to break this stick into infinite pieces. We first sample a value $\psi_1$ from the beta distribution $\mathrm{Beta}(1, \alpha)$ $(\alpha>0)$ and cut the stick at point $\psi_1$ away from the stick's left endpoint. Accordingly, the piece of length $\pi_1(:=\psi_1)$ is retained, and we continue to break the remaining stick with length $1-\pi_1$. Once again, we generate a value $\psi_2$ from $\mathrm{Beta}(1, \alpha)$, cut off $\psi_2$ proportion of the remaining length $1-\pi_1$, and obtain a new piece with length $\pi_2:=(1-\pi_1)\psi_2$. Repeating the breaking procedure on the stick, we have an infinite number of pieces with the $k^{\mathrm{th}}$ piece's length $\pi_k := (1-\sum_{i=1}^{k-1}\pi_{i})\cdot\psi_k$ ($\psi_k \sim \mathrm{Beta}(1, \alpha)$). Each piece $k$ is further given a mark (parameter) $\mu_k$ sampled from a distribution $H$. In this way, we construct a probability measure, $P=\sum_{k=1}^{\infty}\pi_k\delta_{\mu_k}$ ($\delta_{\mu}$ indicates the Dirac measure at $\mu$), with infinite weights $\{\pi_k\}_{k=1}^{\infty}$ and the support on infinite atoms $\{\mu_k\}_{k=1}^{\infty}$. The measure $P$ is said to be from a DP with concentration parameter $\alpha$ and the base distribution $H$, written as $P\sim \mathrm{DP}(\alpha, H)$. Under $P$, each cell $i$ has the probability $\pi_k$ to be from cell type $k$ for any positive integer $k$ without a constraint $K$.
	
	\subsection{Hierarchical Dirichlet process and nested Dirichlet process}
	The DP is only applicable for one level clustering. When another subject level exists, the HDP \citep{hdp} aims to cluster cells for each subject and is able to match cell types in different subjects. In other words, if the cell type indicators $C^{(j_1)}_{i_1}$ and $C^{(j_2)}_{i_2}$ are equal ($j_1$ may not be $j_2$), then the cell $i_1$ in subject $j_1$ and the cell $i_2$ in subject $j_2$ must be from the same cell type. Assume $G^{(j)}$ is the subject-$j$-specific distribution having the form $\sum_{k=1}^{\infty}\pi^{(j)}_{k}\delta_{\mu_k^{(j)}}$, based on which the cells in subject $j$ are clustered. To encourage a common support set across $G^{(j)}$'s, the HDP adopts a hierarchy structure. At the higher level $G_0 \sim \mathrm{DP}(\alpha, H)$, and then at the lower level $G^{(j)}$'s are independent and identically distributed (i.i.d.) and generated from $\mathrm{DP}(\gamma, G_0)$. As the $G_0$ from $\mathrm{DP}(\alpha, H)$ is a discrete distribution and plays the role of the base distribution in $\mathrm{DP}(\gamma, G_0)$, the atoms $\mu^{(j)}_k$'s of the support of $G^{(j)}$ must be consistent with those of $G_0$. This characteristic guarantees the shared cell types across $G^{(j)}$'s in the HDP.
	
	\begin{figure}
		\centering
		\includegraphics[scale = 0.57]{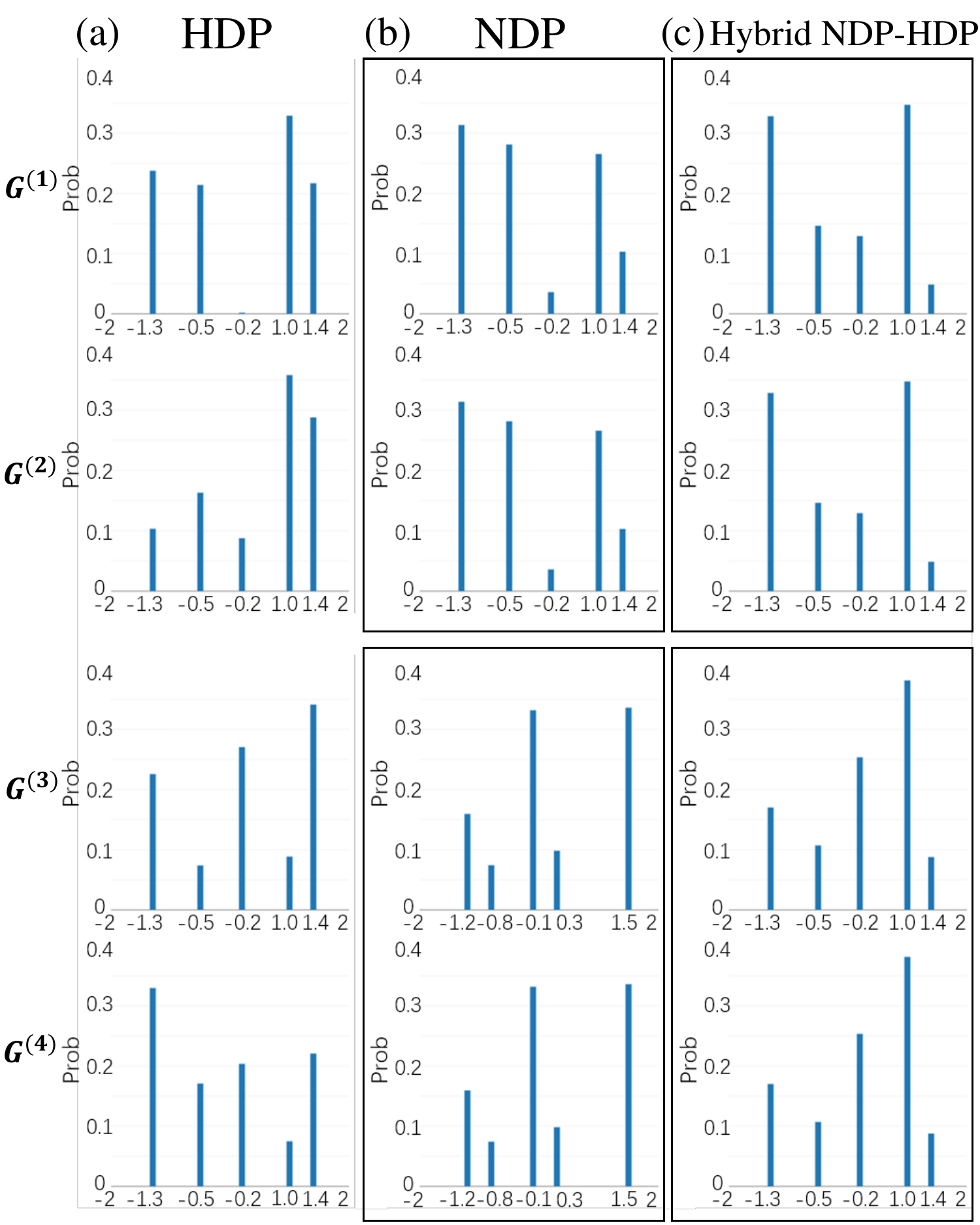}
		\caption{A simple demonstration of three nonparametric Bayesian priors: the HDP, the NDP, and the hybrid NDP-HDP prior. (a) The HDP can make subject-specific distributions $G^{(1)}$, $G^{(2)}$, $G^{(3)}$ and $G^{(4)}$ share the distribution support. However, each distribution $G^{(k)}$ has completely different bar heights (weights) from another. (b) The NDP can achieve the subject subgroup structures; however, two distributions in different subgroups do not have the same support, making it hard to match cell types across subgroups. (c) The hybrid NDP-HDP prior not only groups subject-specific distributions but also enables cell-type-matching between any two subject subgroups. }\label{fig2}
	\end{figure}
	
	Nevertheless, in the HDP, any two subjects have distinct cell distributions due to different weights (cell proportions), i.e., $\mathbb{P}(G^{(j_1)} = G^{(j_2)}) = 0$ if $j_1 \neq j_2$, thus no group structure exists among subjects (Figure \ref{fig2}(a)). The NDP \citep{rodriguez2008nested} permits subject grouping while clustering cells. This ability of the NDP is achieved by replacing the base measure $G_0$ in $\mathrm{DP}(\gamma, G_0)$ with a Dirichlet process $\mathrm{DP}(\alpha, H)$, written as $\mathrm{DP}(\gamma, \mathrm{DP}(\alpha, H))$. Specifically, if we let $Q = \mathrm{DP}(\gamma, \mathrm{DP}(\alpha, H))$, $Q$ takes the form of $\sum_{k=1}^{\infty} \phi_{k}\delta_{G^{\ast}_{k}}$, where the atoms of $Q$ are not numerical values but distributions $G^{\ast}_k$'s from $\mathrm{DP}(\alpha, H)$. Subsequently, $G^{(j)}$'s are i.i.d. sampled from $Q$ and $\mathbb{P}(G^{(j)} = G^{\ast}_k) = \phi_k$. \citet{rodriguez2008nested} showed that there is a positive probability that two distributions $G^{(j_1)}$ and $G^{(j_2)}$ are identical, thus inducing group structures for $G^{(j)}$'s (Figure \ref{fig2}(b)). Despite the simultaneous clustering on subjects and cells enjoyed by the NDP, its assumed continuous measure $H$ leads to totally distinct supports between two subject subgroups (Figure \ref{fig2}(b)). The distributions of the two subjects from the NDP either share all atoms in the support and cell proportions or lack any common atom. Specifically, if $G^{(j_1)}$ and $G^{(j_2)}$ from the NDP have one shared atom, then the whole distribution $G^{(j_1)}$ is equal to $G^{(j_2)}$ almost surely. This is called the degeneracy issue of the NDP outlined in \citet{camerlenghi2019latent}, which causes the difficulty of cell-type-matching for two different subject subgroups in our study.

	\section{The SCSC model}
	The hybrid NDP-HDP prior proposed by \citet{james2008discussion} succeeds in promoting subject subgroups with shared cell types. The nonparametric prior is constructed by assigning a DP prior to the base measure in the NDP,

	{\setlength\abovedisplayskip{-10pt}
		\setlength\belowdisplayskip{-10pt}
		\begin{align}
			G_0 &\sim \mathrm{DP}(\alpha, H), \nonumber\\
			G^{(j)} &\overset{i.i.d.}{\sim} \mathrm{DP}(\nu,\mathrm{DP}(\gamma, G_0)),\;\;\;j=1,\ldots,m. \label{np_prior}
	\end{align}}
	
	\begin{table}
		\caption{Comparing the capabilities of the HDP, the NDP and the hybrid NDP-HDP prior.}
		\label{table1}
		\vskip 0.15in
		\begin{center}
			\begin{small}
				\begin{tabular}{lcc}\hline
					Prior & Subject subgroup structures & Shared support \\\hline
					HDP          & $\times$ &  $\surd$ \\
					NDP          & $\surd$  &  $\times$\\
					Hybrid NDP-HDP prior   & $\surd$  &  $\surd$ \\\hline
				\end{tabular}
			\end{small}
		\end{center}
		\vskip -0.1in
	\end{table}

	On one hand, as $G_0$ is drawn from $\mathrm{DP}(\alpha, H)$, it has a countable support set. This property of $G_0$ makes the child distributions $G^{(j)}$'s share the same support, thus enabling cell-type matching across subjects, an important aspect the NDP lacks. On the other hand, given $G_0$, the NDP helps to form subgroups for subjects. Therefore, the hierarchical and nested nonparametric prior (\ref{np_prior}) integrates the strengths of the HDP and the NDP (Figure \ref{fig2}(c) and Table \ref{table1}).
	
	For the nonparametric prior (\ref{np_prior}), we assume the base measure $H$ is a non-atomic probability measure on measurable space $(U, \mathcal{B})$, where $U$ is a $D$-dimensional subset of $\mathbb{R}^D$ ($U \subset \mathbb{R}^D$), $H(\{y\})=0$ for any $y \in U$, and $\mathcal{B}$ is the Borel $\sigma$-field of $U$. Denote the correlation matrix of the distribution $H$ by $\mathbf{R}_H$. We then have the following results for distributions $G^{(j)}$'s from the prior (\ref{np_prior}).
		
		\begin{proposition} For any Borel set $A \in \mathcal{B}$, we have
			\begin{itemize}
				\item[(1)] $\mathbb{E}\left(G^{(j)}(A)|H\right) = H(A)$.
				\item[(2)] $\mathbb{V}\left(G^{(j)}(A)|H\right) = \frac{\left(\alpha + \gamma + 1\right)H(A)\left(1-H(A)\right)}{\left(\alpha + 1\right)\left(\gamma + 1\right)}$.
				\item[(3)] $\mathrm{Cor}\left(G^{(j)}(A),G^{(j')}(A)|H\right) = \frac{1}{1+\nu}\frac{\nu \gamma + \alpha + \gamma + \nu + 1}{ \alpha + \gamma + 1}$ for $j \neq j'$.
				\item[(4)] When $D=1$, let $\mu_i^{(j)}$ and $\mu_{i'}^{(j')}$ denote random variables from $G^{(j)}$ and $G^{(j')}$, respectively. The correlation between $\mu_i^{(j)}$ and $\mu_{i'}^{(j')}$ is
				\begin{align}
					\mathrm{Cor}\left(\mu_i^{(j)},\mu_{i'}^{(j')}\right) = \left\{ \begin{array}{ll}\frac{\alpha + \gamma + 1}{(\alpha + 1)(\gamma + 1)}&\mathrm{{\it for}}\;j = j',\;i\neq i'\\ \frac{\nu \gamma + \alpha + \gamma + \nu + 1}{ (\nu + 1)(\alpha + 1)(\gamma + 1)}&\mathrm{{\it for}}\;j \neq j'
					\end{array}\right. \nonumber .
				\end{align}
				\item[(5)] When $D \geq 2$, let $\boldsymbol{\mu}_i^{(j)}$ and $\boldsymbol{\mu}_{i'}^{(j')}$ denote the random vectors from $G^{(j)}$ and $G^{(j')}$, respectively. The correlation matrix between $\boldsymbol{\mu}_i^{(j)}$ and $\boldsymbol{\mu}_i^{(j')}$ is
				\begin{align}
					\mathrm{Cor}\left(\boldsymbol\mu_i^{(j)},\boldsymbol\mu_{i'}^{(j')}\right) = \left\{ \begin{array}{ll}\frac{\alpha + \gamma + 1}{(\alpha + 1)(\gamma + 1)}\mathbf{R}_H&\mathrm{{\it for}}\;j = j',\;i\neq i'\\ \frac{\nu \gamma + \alpha + \gamma + \nu + 1}{ (\nu + 1)(\alpha + 1)(\gamma + 1)}\mathbf{R}_H&\mathrm{{\it for}}\;j \neq j'
					\end{array}\right. \nonumber .
				\end{align}
			\end{itemize}
	\end{proposition}
	
	 We notice that when $\alpha$ goes to infinity, $G_0$ in the hybrid NDP-HDP prior approaches to its centering measure $H$. In this limiting case $\alpha \rightarrow +\infty$, the hybrid prior degenerates to the NDP, so the results above are consistent with those in the NDP \citep{rodriguez2008nested}. The proof of the proposition can be found in Supplementary Section S1.
	
	Further, we tailor a zero-inflated distribution to the scRNA-seq raw count data and connect the data-modeling part to the hybrid NDP-HDP prior. One important feature of the scRNA-seq count data is that there is a  relatively  large  proportion of zeros compared to bulk RNA-seq data. This zero-inflation phenomenon, also called dropouts, is mainly caused by a low amount of mRNA molecules in one cell, so the expression levels on some genes are hard to surpass the measurable threshold of the sequencing technology, thus leading to the zero observations.
	
	To model dropout events, we assume that $Y^{(j)}_{gi}$ is the underlying true read count mapped to gene $g$ in cell $i$ for subject $j$; however, these $Y^{(j)}_{gi}$'s are only partially observed through the collected data $X^{(j)}_{gi}$'s due to dropout. As the probability of a dropout happening relies on the value of $Y^{(j)}_{gi}$'s, (i.e., the larger the $Y^{(j)}_{gi}$ the less likely we observe a zero value) the dropout mechanism is ``nonignorable,'' according to the terminology in the field of missing data analysis,
	
	{\setlength\abovedisplayskip{-6pt}
		\setlength\belowdisplayskip{-3pt}
		\begin{align}
			X^{(j)}_{gi} = \left\{ \begin{array}{ll}0&\;\mathrm{with\;probability}\;p(Y^{(j)}_{gi})\\ Y^{(j)}_{gi}&\;\mathrm{with\;probability}\;1-p(Y^{(j)}_{gi}).
			\end{array}\right.\nonumber
		\end{align}
	}
	
	The dropout rate $p(y)$ is modeled as $\Phi(\lambda_{g0} + \lambda_{g1}\log_2(y+1))$ via a Probit link, in which $\lambda_{g1} < 0$ and $\Phi$ is the cumulative distribution function of the standard normal distribution. A negative $\lambda_{g1}$ guarantees negative correlation between $y$ and $p(y)$, and its dependence on the gene index $g$ accurately models the biological observation that the dropout rate may be associated with the gene's features, such as gene length \citep{liu2019hierarchical}.
	
	Due to the count nature and over-dispersion of scRNA-seq data, we adopt the Poisson-log-normal (PLN) distribution for the variable $Y^{(j)}_{gi}$. The PLN distribution has two parameters, $\eta$ and $\sigma^2$, corresponding to the mean and variance of the logarithmic Poisson rate, respectively. Mathematically, $Y \sim \mathrm{PLN}(\eta, \sigma^2)$ if and only if $Y \sim \mathrm{Poi}(e^{\theta})$ and $\theta \sim \mathrm{N}(\eta, \sigma^2)$. This equivalence implies that PLN accounts for the over-dispersion (Supplementary Section S2).
	
	Moreover, one technical factor that can bias the analysis of sequencing data is the {\it library size}, which differs from one cell to another and is defined as the total number of mapped reads to that cell (a detailed description of the library size is given in Supplementary Section S3 and Supplementary Figure S1). To consider the effect of cells' different library sizes, we model $Y^{(j)}_{gi}$ using $Y^{(j)}_{gi} \sim \mathrm{Poi}(s^{(j)}_ie^{\theta^{(j)}_{gi}})$ and $\theta^{(j)}_{gi} \sim \mathrm{N}(\eta^{(j)}_{gi}, \sigma_g^2)$, written as $Y^{(j)}_{gi} \sim \mathrm{PLN}(s^{(j)}_i, \eta^{(j)}_{gi}, \sigma_g^2)$ for simplicity, where  $s^{(j)}_i$ is a scaling factor to consider different library sizes of cells. Specifically, if we denote the library size of cell $i$ in subject $j$ by $l^{(j)}_i$, $s^{(j)}_i$ is calculated as $l^{(j)}_i / median_{i}\; l^{(j)}_i$ and $l^{(j)}_i = \sum_{g=1}^GX^{(j)}_{gi}$ based on the definition of the library size. The $\eta^{(j)}_{gi}$ represents the effects on gene $g$ caused by cell $i$ and subject $j$, and $\sigma_g^2$ reflects variation. We separate cell effects from subject effects and let $\eta^{(j)}_{gi}$ be the addition of the cell-specific effect $\mu^{(j)}_{gi}$ and subject-specific effect $\beta^{(j)}_g$.
	
	Combining the dropout mechanism and the PLN distribution for $Y^{(j)}_{gi}$'s gives the zero-inflated PLN (ZIPLN) distribution for the observed data $X^{(j)}_{gi}$'s, which can be expressed as $X^{(j)}_{gi} \sim \mathrm{ZIPLN}(\lambda_{g0}, \lambda_{g1}, s^{(j)}_i, \mu^{(j)}_{gi} + \beta^{(j)}_g, \sigma_g^2)$. Finally, we assign the nonparametric prior (\ref{np_prior}) to the cell-specific effect vector $\boldsymbol\mu^{(j)}_{i} = (\mu^{(j)}_{1i}, \ldots, \mu^{(j)}_{Gi})^{\top}$ and arrive at the following SCSC model,
	
	
	{\setlength\abovedisplayskip{-8pt}
		\setlength\belowdisplayskip{-5pt}
		\begin{align}
			G_0 &\sim \mathrm{DP}(\alpha, H), \nonumber\\
			G^{(j)} &\overset{i.i.d.}{\sim} \mathrm{DP}(\nu,\mathrm{DP}(\gamma, G_0)),  \;\;\; j=1,\ldots,m,\nonumber \\
			\boldsymbol\mu^{(j)}_{i} &\overset{i.i.d.}{\sim} G^{(j)}, \;\;\; i=1,\ldots,n_j \;\;\;\mathrm{for}\;\mathrm{each}\;j, \nonumber \\
			X^{(j)}_{gi} &\sim \mathrm{ZIPLN}(\lambda_{g0}, \lambda_{g1}, s^{(j)}_i, \mu^{(j)}_{gi} + \beta^{(j)}_g, \sigma_g^2)\;\;\;\mathrm{for}\;\mathrm{each}\;j,\;i,\;\mathrm{and}\;g.\label{scsc}
		\end{align}
	}

	Here, the base measure $H$ is a non-atomic probability measure on the measurable space $(\mathbb{R}^D, \mathcal{B})$, where $\mathbb{R}^D$ is a real coordinate space of dimension $D$ and $\mathcal{B}$ is the Borel $\sigma$-field of $\mathbb{R}^D$. We constrain the subject-specific effects $\beta^{(j_1)}_g = \beta^{(j_2)}_g$ for any $g$ if $G^{(j_1)} = G^{(j_2)}$, as subjects from the same subgroup usually exhibit the same characteristic. Moreover, to make the parameters $\boldsymbol\mu$ and $\boldsymbol\beta$ estimable, we let one subject subgroup act as the ``reference'' group and constrain the subject effects $\boldsymbol\beta^{(j)}$ of the reference group to be zero.

	\section{The truncated SCSC model}
	Exact posterior sampling for the SCSC model can be performed by the Polya-urn scheme \citep{Pitman1996Some}, which marginalizes the distributions $G_0$ and $G^{(j)}$'s $(j\geq 1)$. However, the marginalization procedure introduces extra dependence among cells and causes the cell-type allocation update for one cell to rely on all other cells. Such a sequential update scheme results in unnecessary and heavy computations. Therefore, to enhance posterior sampling efficiency for the SCSC model, we utilize the blocked Gibbs sampler \citep{ishwaran2001gibbs}, where the updates in each parameter block are independent, by taking a truncation strategy \citep{ishwaran2001gibbs, rodriguez2008nested}---setting the upper bounds $L$ for the number of subject subgroups and $K$ for the cell type number. Moreover, the blocked Gibbs sampler also favors the use of parallel computing to further speed up posterior inference. The truncated SCSC model is

	{\setlength\abovedisplayskip{-8pt}
		\setlength\belowdisplayskip{-5pt}
		\begin{align}
			G_0 &\sim \mathrm{DP}(\alpha, H), \nonumber\\
			G^{(j)} &\overset{i.i.d.}{\sim} \mathrm{DP}_{L}(\nu,\mathrm{DP}_{K}(\gamma, G_0)),\;\;\; j=1,\ldots,m , \nonumber \\
			\boldsymbol\mu^{(j)}_{i} &\overset{i.i.d.}{\sim} G^{(j)},\;\;\; i=1,\ldots,n_j \;\;\;\mathrm{for}\;\mathrm{each}\;j,\nonumber \\
			X^{(j)}_{gi} &\sim \mathrm{ZIPLN}(\lambda_{g0}, \lambda_{g1}, s^{(j)}_i, \mu^{(j)}_{gi} + \beta^{(j)}_g, \sigma_g^2)\;\;\;\mathrm{for}\;\mathrm{each}\;j,\;i,\;\mathrm{and}\;g.\label{scsc_trun}
		\end{align}
	}

	Using the stick-breaking process metaphor, $\mathrm{DP}_{K}(\gamma, G_0)$ indicates that we break the unit stick into $K$ pieces rather than infinite pieces. The following theorem states that the truncation model (\ref{scsc_trun}) is an accurate approximation to the original model (\ref{scsc}) as long as the concentration parameters $\gamma$ and $\nu$ as well as the truncation numbers $L$ and $K$ are appropriately selected. The choice of $(\nu, \gamma, K, L)$ is discussed later. See the Supplementary Section S4 for the proof, which is based on the Theorem B1 in the NDP paper \citep{rodriguez2008nested}.
	
	\begin{theorem}
		Denote the prior distributions of cell effects $\boldsymbol\mu$ from the SCSC model and the truncated SCSC model by $p^{\infty\infty}(\boldsymbol\mu)$ and $p^{KL}(\boldsymbol\mu)$, respectively. Based on the priors, we have the marginal distributions $p^{\infty\infty}(\mathbf{x})$ and $p^{KL}(\mathbf{x})$ for the observed data $\mathbf{x}$ by integrating all parameters out. We then have
		\begin{align}
			\frac{1}{4} \int \left|p^{KL}(\mathbf{x})-p^{\infty\infty}(\mathbf{x})\right| d\mathbf{x} \leq 1 - \left\{1 - \left(\frac{\nu}{\nu+1}\right)^{L-1}
			\right\}^m\left\{1 - \left(\frac{\gamma}{\gamma + 1}\right)^{K-1} \right\}^{\sum\limits_{j=1}^mn_j}. \nonumber
		\end{align}
	\end{theorem}

	If we expand the implicit distributions $G^{(j)}$'s in model (\ref{scsc_trun}) in terms of subject cluster indicators $S^{(j)}$'s and cell type indicators $C^{(j)}_i$'s, then we obtain a more concrete and interpretable model.
	
	{\setlength\abovedisplayskip{-8pt}
		\setlength\belowdisplayskip{-5pt}
		\begin{align}
			\boldsymbol\xi = (\xi_1, \xi_2, \ldots, \xi_K) &\sim \mathrm{GEM}_{K}(\alpha),\nonumber \\
			\boldsymbol\mu_{k} &\overset{i.i.d.}{\sim} H, \;\;\; k=1,\ldots, K, \nonumber\\
			\boldsymbol\pi_{\ell} = (\pi_{1\ell}, \ldots, \pi_{K\ell}) &\overset{i.i.d.}{\sim} \mathrm{Dir}(\gamma\xi_1, \gamma\xi_2, \ldots, \gamma\xi_K),\;\;\; \ell=1,\ldots,L, \nonumber \\
			\boldsymbol\phi = (\phi_1, \phi_2, \ldots, \phi_L) &\sim \mathrm{GEM}_{L}(\nu), \nonumber\\
			S^{(j)} &\overset{i.i.d.}{\sim} \mathrm{MN}(1;\phi_1, \phi_2, \ldots, \phi_L),\;\;\;j=1,\ldots,m,  \nonumber \\
			C^{(j)}_i | S^{(j)}=\ell &\overset{i.i.d.}{\sim} \mathrm{MN}(1;\pi_{1\ell}, \ldots, \pi_{K\ell}),\;\;\;i=1,\ldots,n_j \;\;\;\mathrm{for}\;\mathrm{each}\;j,\nonumber \\
			X^{(j)}_{gi}| S^{(j)}=\ell, C^{(j)}_i=k  &\sim \mathrm{ZIPLN}(\lambda_{g0}, \lambda_{g1}, s^{(j)}_i, \mu_{gk} + \beta_{g\ell}, \sigma_g^2)\;\;\;\mathrm{for}\;\mathrm{each}\;j,\;i,\;\mathrm{and}\;g.\label{scsc_sim}
		\end{align}
	} $\mathrm{MN}$ is the multinomial distribution and $\mathrm{Dir}$ indicates the Dirichlet distribution. $\mathrm{GEM}_{L}(\nu)$ refers to the truncated stick-breaking process in which the stick proportions $\{\phi'_1, \phi'_2, \ldots, \phi'_{L-1}\}$ are i.i.d. from $\mathrm{Beta}(1, \nu)$ and $\phi_{1} = \phi'_1$, $\phi_{\ell} = \phi'_{\ell}\prod_{t=1}^{\ell-1}(1-\phi'_{t})$ for $2\leq \ell \leq L-1$, and $\phi_{L} = 1 - \sum_{\ell=1}^{L-1}\phi_{\ell}$. This is similar for $\mathrm{GEM}_{K}(\alpha)$. Again, we note that the subgroup one effect vector $\boldsymbol\beta_{1}$ is fixed at zero for identifiability. We prove that model (\ref{scsc_sim}) is equivalent to model (\ref{scsc_trun}) in Supplementary Section S5. Subsequently, we focus on model (\ref{scsc_sim}) to perform the Bayesian inference.
	
	We note that in the stick-breaking process the prior expectation of the first stick's length is always larger than others and in practice we usually assign the first subgroup as the reference group, so we need to be cautious about the choice of $\nu$ that reflects our prior belief for the relative weight of the reference group ($1/(1+\nu)$ in expectation). If we replaced the truncated stick-breaking prior in Model (\ref{scsc_sim}) $\boldsymbol\phi = (\phi_1, \phi_2, \ldots, \phi_L) \sim \mathrm{GEM}_{L}(\nu)$ by a finite dimensional Dirichlet prior \citep{ishwaran2001gibbs} $\boldsymbol\phi = (\phi_1, \phi_2, \ldots, \phi_L) \sim \mathrm{Dir}(\nu/L, \nu/L, \ldots, \nu/L)$, it would mitigate the effect of the prior weight bias induced by the truncated stick-breaking process, but this replacement breaks the equivalence between Models (\ref{scsc_trun}) and (\ref{scsc_sim}).
	
	In Model (\ref{scsc_sim}), a larger $\nu$ encourages more subject subgroups, and a larger $\gamma$ reflects that the cell proportions across subject subgroups have more concentration on the normalized $(\xi_1, \xi_2, \ldots, \xi_K)$ whose assignments are determined by $\alpha$. Thus, we first choose $\gamma$ and $\nu$ to reflect our prior belief and then chosse $K$ and $L$ appropriately to guarantee a small approximation error. Throughout the paper, we chose $\nu=\gamma=0.5$ and $K=L=15$, giving a small approximation error in simulation and real application.
	
	As we cluster high-dimensional expression data, it is important to conduct feature selection. \citet{tadesse2005bayesian} proposed a Bayesian variable selection method to cluster high-dimensional samples and identify discriminating variables simultaneously, so we incorporated this idea into the proposed model SCSC, resulting in a variable selection version, which we termed SCSC-vs. Details can be found in Supplementary Section S6.
	
	\section{Bayesian posterior inference}
	We next specify the priors for unknown parameters in model (\ref{scsc_sim}). The prior for concentration parameter $\alpha$ $(\alpha>0)$ is a gamma distribution, $\alpha \sim \mathrm{\Gamma}(a_{\alpha_1}, a_{\alpha_2})$. Regarding the baseline distribution $H$ of cell-type-$k$ effects $\mu_{gk}$'s, it is set as the Cartesian product of $D$ normal distributions $\mathrm{N}(\eta_{\mu}, \tau^2_{\mu})$, and we assign hyper-priors $\eta_{\mu} \sim \mathrm{N}(u_{\mu}, \omega^2_{\mu})$ and $\tau^2_{\mu} \sim \mathrm{Inv\Gamma}(b_{\mu1}, b_{\mu2})$ to $\eta_{\mu}$ and $\tau^2_{\mu}$, respectively. Similarly, we assign a normal distribution $\mathrm{N}(\eta_{\beta}, \tau^2_{\beta})$ to the subgroup effect $\beta_{g\ell}$ and further assign $\eta_{\beta}$ and $\tau^2_{\beta}$ hyper-priors $\eta_{\beta}  \sim \mathrm{N}(u_{\beta}, \omega^2_{\beta})$ and $\tau^2_{\beta} \sim \mathrm{Inv\Gamma}(b_{\beta1}, b_{\beta2})$ to introduce hierarchy for subject effects. This enables information to be borrowed across genes. The variance $\sigma_g^2$'s prior distribution is an inverse-gamma distribution $\sigma_g^2 \sim \mathrm{Inv\Gamma}(b_{\sigma1}, b_{\sigma2})$, and the priors for zero-inflation-related parameters $\lambda_{g0}$ and $\lambda_{g1}$ are given by weakly informative priors $\mathrm{N}(\eta_{\lambda_{g0}}, \tau^2_{\lambda_{g0}})$ and $\mathrm{N}(\eta_{\lambda_{g1}}, \tau^2_{\lambda_{g1}})\mathbb{I}(\lambda_{g1} < 0),$ respectively.
	
	Finally, given the priors and model (\ref{scsc_sim}), we utilize the blocked Gibbs sampler \citep{ishwaran2001gibbs} to perform the posterior sampling. As directly sampling from ZIPLN distribution suffers from an intractable infinite sum and integral, we augment the model with the auxiliary variables $\theta^{(j)}_{gi}$ and $Y^{(j)}_{gi}$\citep{Tanner1987The} specified in Section 3 to make the sampling for ZIPLN feasible. The Gibbs sampling scheme is presented in detail in Supplementary Section S7.
	Some steps of the blocked Gibbs sampler do not correspond to tractable distributions; hence we adopt a Metropolis-within-Gibbs framework in such cases. The proposal distributions and the calculations of acceptance rates are contained in Supplementary Section S8. For each iteration of the Gibbs sampler, the computational complexity is $\mathcal{O}(DKL\sum_{j=1}^mn_j)$, which increases linearly with the gene number $D$, the total cell number $\sum_{j=1}^mn_j$, and the upper bounds $K, L$. Thus, the MCMC algorithm can scale well on a large volume of scRNA-seq data.
	
	
	After the burn-in period which is defined as the first half of the iterations, we collect the posterior samples from the last half of iterations for statistical inference. Further, we estimate the subgroup and cell-type indicators, $S^{(j)}$'s and $C^{(j)}_i$'s, using the mode of the posterior samples to keep the integer nature. For the subgroup effects and cell-type-specific effects, $\beta_{g\ell}$'s and $\mu_{gk}$'s, respectively, the posterior mean is used for estimation.

	\section{Results}
	\subsection{Simulation}
\begin{figure}
		\centering
		\includegraphics[scale = 0.37]{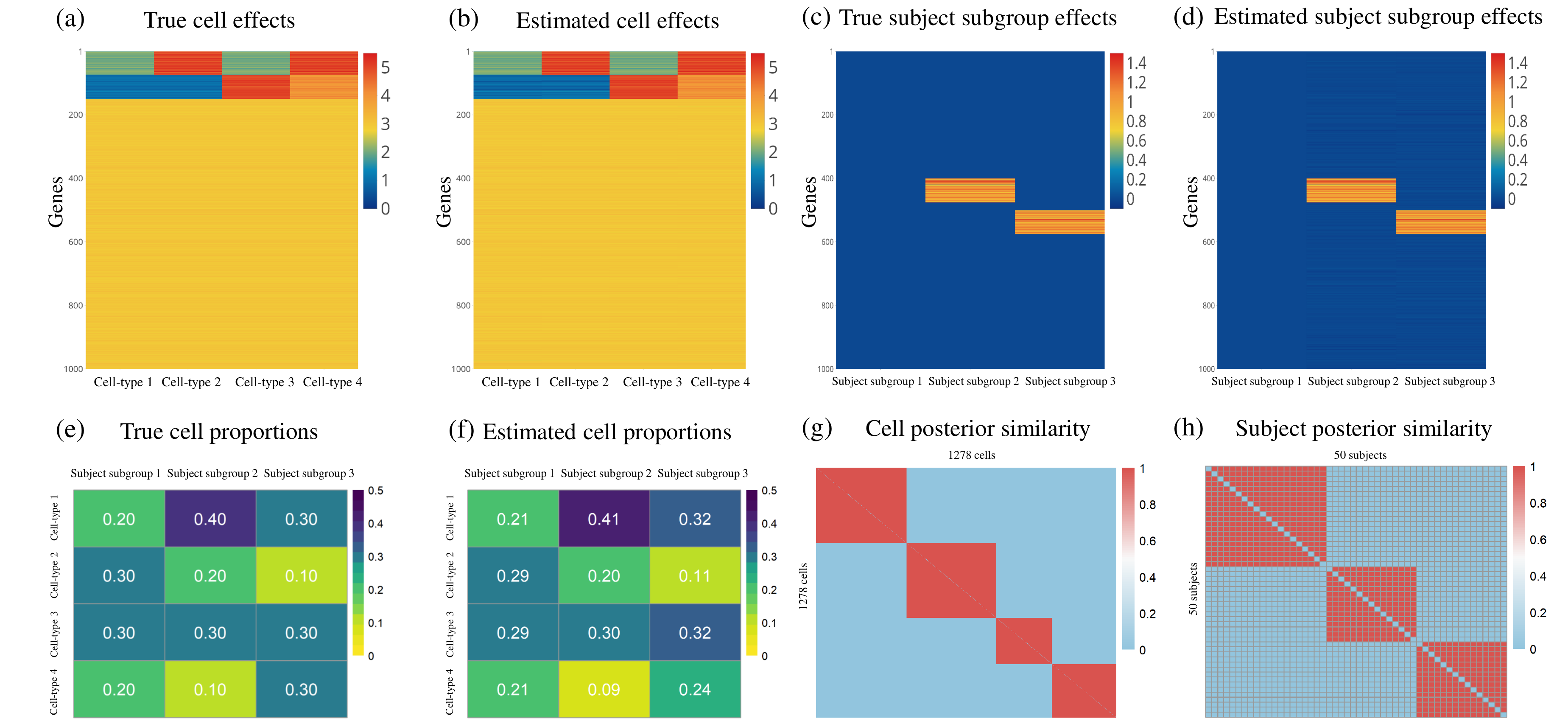}
		\caption{Performance of the SCSC model in the simulation. (a) Heatmap of the true cell effects $\mu_{gk}$ and (b) heatmap of cell effect estimations. In both (a) and (b), one row represents one gene, and each column represents one cell type. (c) Heatmap of the true subject subgroup effects and (d) heatmap of subject subgroup effect estimations. In both (c) and (d), one row represents one gene and each column represents one subject subgroup. (d) Heatmap of the true cell proportions for each subgroup. (e) The cell proportion estimates. (g-h) Posterior similarity matrix heatmaps for (g) cells and (h) subjects. In the similarity matrix, the $(i, j)$ element is the posterior probability that objects $i$ and $j$ are in the same cluster for $i\neq j$.}\label{fig3}
	\end{figure}

	We generated data following model (\ref{scsc_sim}) with generation details provided in Supplementary Section S9. We then applied our SCSC model to this dataset using $\gamma = \nu = 0.5$, the subject subgroup upper bound $L=15$, and cell type number upper bound $K=15$, which guarantees a small approximation error $0.0011$ based on Theorem 1. We performed $10,000$ iterations. By correcting the label-switching (Supplementary Section S10), we evaluated the estimates of SCSC for the cell type effects $\boldsymbol\mu$, subgroup effects $\boldsymbol\beta$, and cellular proportions for each subject subgroup $\boldsymbol\pi$. The comparison between the true parameter values and the estimates is shown in Figure \ref{fig3}(a-f), indicating that the SCSC model estimated these parameters well. Figure 3(g-h) displays the posterior similarity matrices for cell clustering and subject clustering, respectively, showing clear clustering structures for cells and subjects. Hence, the SCSC model can automatically and accurately distinguish the underlying heterogeneity for subjects and cells.
	
	\begin{figure}
		\centering
		\includegraphics[scale = 0.65]{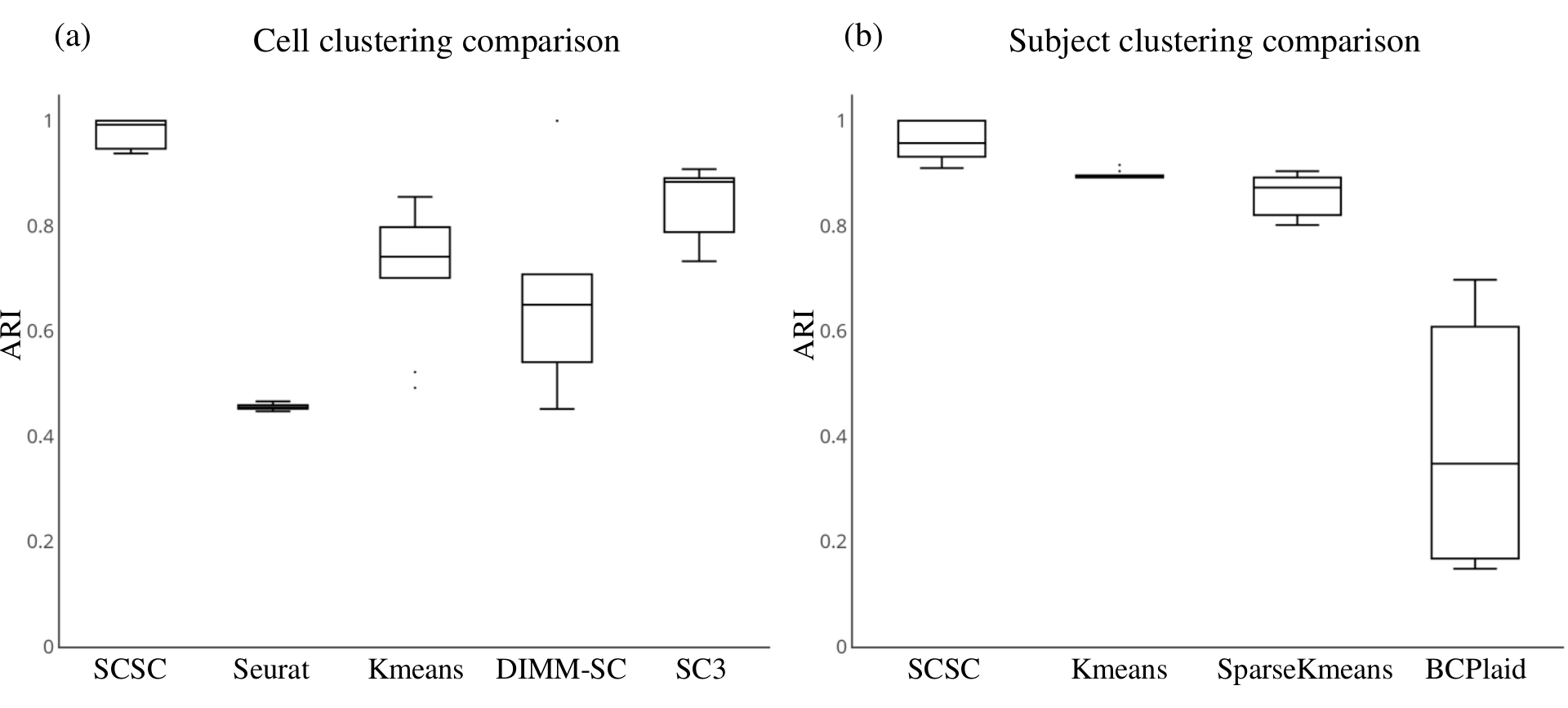}
		\caption{Clustering performances of the SCSC model as well as competing methods in the cell clustering and subject clustering settings based on ten realizations. (a) ARI box plots for SCSC and other cell clustering approaches. (b) ARI box plots for SCSC and other subject clustering approaches. The implementation details of the competing methods are provided in Supplementary Section S11.}\label{fig4}
	\end{figure}
	
	As there is no statistical approach to simultaneously cluster subjects and cells, we compared the SCSC against some popular cell and subject clustering approaches, respectively. We selected cell clustering approaches k-means \citep{macqueen1967some}, SC3 \citep{Kiselev2017SC3}, DIMM-SC \citep{Sun2017DIMM}, and Seurat \citep{butler2018integrating, stuart2018comprehensive} and subject clustering approaches kmeans \citep{macqueen1967some}, SparseKmeans \citep{witten2010framework}, and BCPlaid \citep{turner2005improved}.  The boxplots for ARI values \citep{Hubert1985Comparing} of all methods under the cell and subject clustering setting based on ten realizations are shown in Figure \ref{fig4}. Overall, SCSC performed better in both cell clustering and subject clustering. When clustering cells, SCSC borrows information across multiple subjects and considers the subject differences. When grouping subjects, the model exploits the cell information of each subject to discover the subtle difference. Owing to the two-way information-sharing strategy, SCSC outperforms competing methods in both cell clustering and subject grouping.
	
	The performances of SCSC and SCSC-vs on low signal scenarios and model misspecification cases were discussed in Supplementary Section S12.
	
	\subsection{Real application}
	\citet{sarkar2019discovery} collected scRNA-seq datasets from 7,585 induced pluripotent stem cells (iPSCs) in a total of 54 Yoruba subjects in Nigeria. The datasets are publicly available with the accession code GSE118723 in GEO \citep{edgar2002gene}. Although the purpose of the study \citep{sarkar2019discovery} was to detect variance QTLs, we can use the same dataset to mine out other interesting information, such as the cell and subject heterogeneity presented here. At the subject level, Yoruba is one of Nigeria's largest ethnic groups, and the Yorubas in the same lineage are more likely to suffer from the same genetic diseases \citep{olaitan2014recruitment}. Therefore, analyzing the heterogeneity of the Yorubas can clarify their family relationships or find Yoruba sub-races. At the cell level, the iPSCs are reprogrammed from the somatic cells in adult tissues and have the ability to differentiate into several cell types. Hence, they can be potentially used to make personalized treatments for patients. The iPSCs derived from different somatic cell types may demonstrate heterogeneous differentiation abilities \citep{kim2011donor}. Our aim is to apply SCSC to the dataset to distinguish Yoruba individuals and separate the iPSC heterogeneity at the same time.
	
	Our analysis focused on scRNA-seq counts from batch 6 in \citet{sarkar2019discovery}, which includes 20 subjects and 1,152 cells. In the preprocessing procedure, we filtered out cells with the zero proportion of more than 80\% and genes with the zero proportion of more than 30\%. We further removed subjects having less than 5 cells, resulting in a scRNA-seq dataset with 14 subjects, 1,028 cells, and 4,178 genes. The cell numbers of the selected 14 subjects ranged from 29 to 129. During the analysis, the scaling factors were computed to adjust the effects of library sizes.
	
	\begin{figure}
		\centering
		\includegraphics[scale = 0.6]{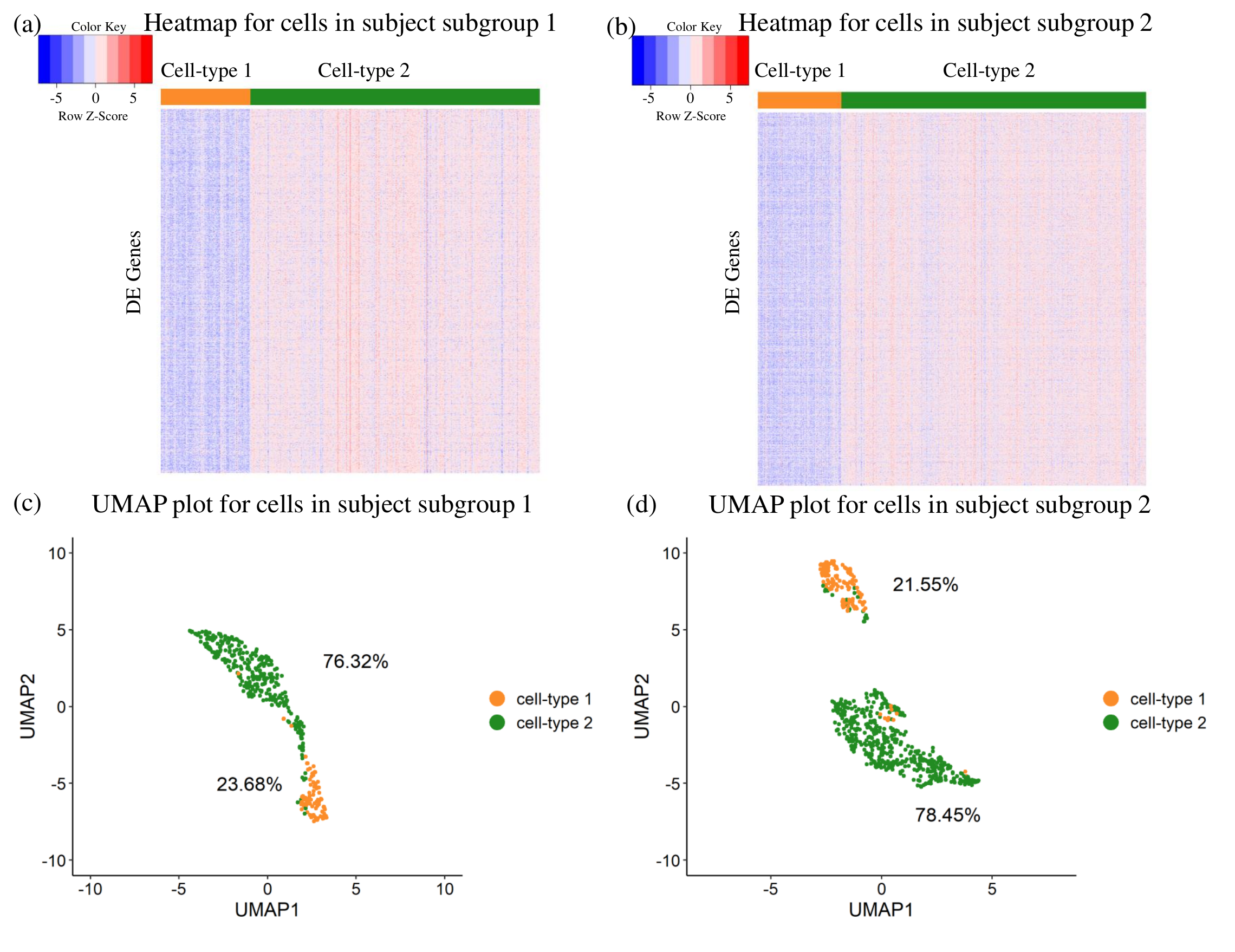}
		\caption{Performance of SCSC on the Yoruba iPSC scRNA-seq data. (a) Heatmap for the logarithm-transformed and row-scaled gene expression values of cells in subject subgroup 1. There are 2,698 DE genes, 94 type 1 cells, and 303 type 2 cells. Cells under the same color are from the same cell type. (b) Heatmap for logarithm-transformed and row-scaled gene expression values of cells in subject subgroup 2. There are 2,698 DE genes, 136 type 1 cells, and 495 type 2 cells. (c-d) Scatter plots by projecting cells in subject subgroups 1 and 2 to a two-dimensional space using UMAP via R package umap \citep{umapRpackage}. Cells are colored by the estimated cell types: cell type 1 (orange), cell type 2 (green).}\label{fig5}
	\end{figure}
	
	We then implemented the SCSC model with $(\gamma, \nu, K, L)$ $=$ $(0.5, 0.5, 15, 15)$, resulting in a small approximation error $0.0009$. The blocked Gibbs sampler performed 10,000 iterations with the first half as the burn-in period, and it took about 21.66 hours using 24 CPU cores. The trace plots in Supplementary Figure S2 showed that the chains had attained convergence during burn-in.  Two Yoruba subgroups and two iPSC types were identified. The posterior similarity matrix heatmaps for cells and subjects are in Supplementary Figure S3. Yoruba subgroup 1 contained 4 subjects and had cellular compositions of 23.68\%, 76.32\% for cell types 1 and 2, respectively. Yoruba subgroup 2 contained 10 subjects with cell type compositions of 21.55\%, 78.45\%. The heatmaps for logarithm-transformed and row-scaled expression values in Yoruba subgroups 1 and 2, respectively, are shown in Figure \ref{fig5}(a) and \ref{fig5}(b). We observed clearly differential expression patterns between cell types 1 and 2 on detected cell type DE genes, indicating the existence of heterogeneity among iPSCs. Besides the cellular compositions, the estimated effects of the Yoruba subgroups also demonstrated the heterogeneity of the Yoruba individuals (Supplementary Figure S4).  A clear cell pattern in Yoruba subgroups 1 and 2 is observed in Figure \ref{fig5}(c-d): cells of type 1 (orange) and type 2 (green) are well-separated. Sensitivity analyses (Supplementary Section S13 and Supplementary Figures S5-7) demonstrate that the clustering result obtained by SCSC is robust to the choices of hyper-parameters. Moreover, the validation of the SCSC clustering results are provided in Supplementary Section S14.

	\section{Conclusion}
	In this study, we developed a nonparametric Bayesian model, SCSC, to simultaneously discover subject and cell heterogeneity in a two-level clustering approach. SCSC has the flexibility of learning the subject subgroup or cell type number from the data without a prespecification. Unlike priors such as the HDP or the NDP, we employed the hybrid NDP-HDP prior \citep{james2008discussion} to induce group structures in subjects, cluster cells in each subject, and match cell types across subjects. The ZIPLN distribution developed in SCSC directly models the count nature, over-dispersion, and dropouts of the scRNA-seq data. Owing to these two features, the SCSC model achieves the subject-level and cell-level clustering on the multi-subject scRNA-seq data. When clustering subjects, SCSC takes advantage of the cell-resolution differences; when clustering cells, it borrows information across multiple subjects. The two-way information-sharing strategy enables SCSC to obtain more accurate clustering results than competing methods in the domain of either subject clustering using bulk expression data or cell clustering based on scRNA-seq data.
	
	To the best of our knowledge, SCSC is the first unified approach in addressing the two-level clustering for scRNA-seq data. Notably, SCSC bridges the methodology gap between subject clustering based on aggregated gene expression data and scRNA-seq cell clustering. The framework in SCSC can be further adapted to situations where the observed data are sparse, count-valued and two-level clustering are of interest. In the meanwhile, there are some directions we can extend the SCSC model. All distributions induced by the hybrid NDP-HDP prior have the same atoms. However, one subject subgroup may have its own cell-type. For example, one tumor subtype can have its unique tumor cell subclone. The incorporation of semi-HDP \citep{beraha2020semi} can help generate distributions there exist both shared and unique atoms. Additionally, the DP is a special case of the Pitman-Yor process \citep{pitman1997two} that has many desirable features in practice, so replacing the HDP by hierarichical Pitman-Yor processes would create more realistic clustering behavior especially in scRNA-seq data analysis \citep{camerlenghi2019nonparametric}.
	
	Considering the continuous progress of the sequencing technology, single-cell RNA sequencing will be affordable and available for more persons. Therefore, we envision that the SCSC model will be a useful method to facilitate the development of personalized treatment in a time of single-cell genomics.
	
	\section*{Supplementary Materials} Proofs, MCMC derivations, and some results in simulation and real application are in a separate supplementary file (\url{https://drive.google.com/file/d/1svOJbjAjhdN1g0IWOs4o2jEl41WWm2MU/view?usp=sharing}). The R package to implement SCSC is available on GitHub \url{https://github.com/WgitU/SCSC}.
	\par
	
	

	\bibhang=1.7pc
	\bibsep=2pt
	\fontsize{9}{14pt plus.8pt minus .6pt}\selectfont
	\renewcommand\bibname{\large \bf References}
	\expandafter\ifx\csname
	natexlab\endcsname\relax\def\natexlab#1{#1}\fi
	\expandafter\ifx\csname url\endcsname\relax
	\def\url#1{\texttt{#1}}\fi
	\expandafter\ifx\csname urlprefix\endcsname\relax\def\urlprefix{URL}\fi
	
	\bibliographystyle{apalike}
	\bibliography{SCSC}   
	
	\vskip .65cm
	\noindent
	Qiuyu Wu,
	\vskip 2pt
	\noindent
	E-mail: w.qy@ruc.edu.cn
	\vskip 2pt
	
	\noindent
	Xiangyu Luo (corresponding author),
	\vskip 2pt
	\noindent
	E-mail: xiangyuluo@ruc.edu.cn
	
\end{document}